\documentclass[aps,prb,amsmath,amssymb,twocolumn,showpacs,superscriptaddress,final,floatfix]{revtex4-1}
\usepackage{graphicx}	
\usepackage{color}	
\usepackage[normalem]{ulem}
\usepackage{hyperref}
\hypersetup{
  colorlinks,
  citecolor=blue,
  linkcolor=blue,
  urlcolor=blue}

\newcommand{\YRG}{YbRu$_2$Ge$_2$}
\newcommand{\YRS}{YbRh$_2$Si$_2$}

\newcommand{\ie}{{\em i.e.}}
\newcommand{\eg}{{\em e.g.}}

\begin{document}

\title{Electronic structure of the quadrupolar ordered heavy-fermion compound YbRu$_2$Ge$_2$ measured by angle-resolved photoemission}

\author{H. Pfau}
\email{hpfau@stanford.edu}
\affiliation{Stanford Institute for Materials and Energy Sciences, SLAC National Accelerator Laboratory, 2575 Sand Hill Road, Menlo Park, CA 94025, USA}
\affiliation{Department of Physics, Stanford University, Stanford, CA 94305, USA}

\author{E. W. Rosenberg}
\author{P. Massat}
\affiliation{Stanford Institute for Materials and Energy Sciences, SLAC National Accelerator Laboratory, 2575 Sand Hill Road, Menlo Park, CA 94025, USA}
\affiliation{Geballe Laboratory for Advanced Materials, Department of Applied Physics, Stanford University, Stanford, CA 94305, USA}

\author{B. Moritz}
\affiliation{Stanford Institute for Materials and Energy Sciences, SLAC National Accelerator Laboratory, 2575 Sand Hill Road, Menlo Park, CA 94025, USA}

\author{M. Hashimoto}
\author{D. Lu}
\affiliation{Stanford Synchrotron Radiation Lightsource, SLAC National Accelerator Laboratory, 2575 Sand Hill Road,
Menlo Park, California 94025, USA}

\author{I. R. Fisher}
\affiliation{Stanford Institute for Materials and Energy Sciences, SLAC National Accelerator Laboratory, 2575 Sand Hill Road, Menlo Park, CA 94025, USA}
\affiliation{Geballe Laboratory for Advanced Materials, Department of Applied Physics, Stanford University, Stanford, CA 94305, USA}

\author{Z.-X. Shen}
\affiliation{Stanford Institute for Materials and Energy Sciences, SLAC National Accelerator Laboratory, 2575 Sand Hill Road, Menlo Park, CA 94025, USA}
\affiliation{Department of Physics, Stanford University, Stanford, CA 94305, USA}
\affiliation{Geballe Laboratory for Advanced Materials, Department of Applied Physics, Stanford University, Stanford, CA 94305, USA}

\date{\today}

\begin{abstract}
We studied the electronic structure of the heavy fermion compound Yb(Ru$_{1-x}$Rh$_{x}$)$_2$Ge$_2$ with $x=0$ and nominally $x=0.125$ using ARPES and LDA calculations. We find a valence band structure of Yb corresponding to a non-integer valence close to $3+$. The three observed crystal electric field levels with a splitting of 32 and 75 meV confirm the suggested configuration with a quasi-quartet ground state. The experimentally determined band structure of the conduction electrons with predominantly Ru $4d$ character is well reproduced by our calculations. \YRG~undergoes a non-magnetic phase transition into a ferroquadrupolar ordered state below 10.2\,K and then to an antiferromagnetically ordered state below 6.5\,K. A small hole Fermi surface shows nesting features in our calculated band structure and its size determined by ARPES is close to the magnetic ordering wave vector found in neutron scattering.  The transitions are suppressed when \YRG~is doped with 12.5\% Rh. The electron doping leads to a shift of the band structure and successive Lifshitz transitions. 
\end{abstract}

\maketitle

\section{Introduction}

Heavy-fermion systems are an ideal playground to study quantum critical phenomena \cite{loehneysen_2007}. Multipole ordering is an exotic symmetry breaking \cite{morin_rev_1990} that offers the possibility of interesting types of quantum criticality \cite{si_2013,lai_arxiv}. A prominent example for multipole ordering is CeB$_6$ with an antiferroquadrupole ordering.\cite{effantin_1985,kuramoto_2009} It has a cubic symmetry with a quartet crystal electric field (CEF) ground state. Another example is the series of Pr$T_2 X_{20}$, which contains a number of compounds with quadrupolar order.\cite{onimaru_2011,sakai_2011,onimaru_2012} Here, the CEF from the cubic lattice results in a non-magnetic ground state of the Pr $4f^2$ which has a nonzero matrix elements for quadrupoles.  

Of the various types of multipolar order, ferroquadrupole (FQ) order is particularly interesting in light of recent results on nematicity in the Iron-based high-temperature superconductors (FeSC). \cite{fernandes_2014} Both ordering phenomena break rotational but not translational symmetry. While nematicity in the FeSC is driven by an instability of the itinerant Fe $3d$ electrons, FQ order in heavy fermion systems involves local $4f$ electrons. The coupling of orbital, spin and lattice degrees of freedom was explored extensively in the FeSC using for example strain as a tuning parameter for the nematic order. \cite{chu_2012,yi_2011_pnas,beak_2014} 
However, the underlying physics of the nematicity in FeSC is complex with several other intertwined orders. In contrast, the driving force behind FQ order of local $4f$ orbitals is well understood \cite{morin_rev_1990}. Consequently, materials that undergo a FQ phase transition can serve as model systems to study the effects of nematic fluctuations on conduction electrons, motivating detailed study of their electronic properties.

Here we investigate \YRG, which presents a recent example of FQ order. It shows weak Kondo lattice behavior with a Sommerfeld coefficient of $\gamma=100\,\mathrm{mJ/K^2}$. \cite{jeevan_2006} The crystal electric field (CEF) of the tetragonal crystal structure splits the Yb $4f$ levels into four Kramers doublets. However, the ground state is a quasi-quartet with a splitting of less than 1\,meV. \cite{jeevan_2006,jeevan_diss,rosenberg_arxiv} Neutron scattering determined the splitting to the third CEF level to be 32\,meV.\cite{jeevan_2011} The splitting to the fourth level was estimated to be 91\,meV.\cite{jeevan_diss} Three phase transitions were observed at $T_0=10.2$\,K, $T_1=6.5$\,K and $T_2=$5.7\,K. \cite{jeevan_2006} $T_1$ could be ascribed to an antiferromagnetic ordering with an incommensurate wave vector of $q=(0.352,0,0)$.\cite{jeevan_2006,jeevan_2011} A slight change of the propagation vector appears at $T_2$. \cite{jeevan_2011} No signatures of magnetic order were found at $T_0$ in neutron scattering or muon-spin resonance and a theoretical analysis based on the quasi-quartet ground state suggested a FQ order.\cite{jeevan_2006,takimoto_2008} Recent low-temperature x-ray scattering experiments confirm the presence of a tetragonal to orthorhombic phase transition at $T_0$. \cite{rosenberg_arxiv} Measurements of the quadrupole strain susceptibility reveal a Curie behavior demonstrating that the FQ order is primarily driven by magneto-elastic coupling. \cite{rosenberg_arxiv} 

We performed angle-resolved photoelectron spectroscopy (ARPES) and LDA band structure calculations to shed light on the electronic structure of \YRG. We study both the $4f$ and the conduction electron system. Our results confirm the weak heavy fermion behavior with a non-integer Yb valence close to 3+. Three CEF levels can be identified with a splitting of 32(6)\,meV and 75(6)\,meV. The Ru $4d$ conduction bands hybridize with the local $4f$ levels. The $4d$-electron band structure is well reproduced by our calculations and we find a small Fermi surface that contains favorable nesting properties with a nesting vector close to the magnetic ordering vector. Doping \YRG~with Rh suppresses the magnetic and quadrupole order and we detect three phase transition between 2\,K and 3\,K at nominally 12.5\% doping. We observe a shift in the band structure due to the electron doping which results in two Lifshitz transitions: one very close to and one well below 12.5\% Rh doping.


\section{Experimental Methods}

High quality single crystals of \YRG~and Yb(Ru$_{0.875}$Rh$_{0.125}$)$_2$Ge$_2$ were grown in indium flux as described in Ref.~\onlinecite{rosenberg_arxiv}. The residual resistivity ratio is 10 for the parent compound \cite{rosenberg_arxiv}. We characterized the phase transitions with specific heat measurements using a relaxation technique performed in a PPMS from Quantum Design. The ARPES measurements were performed at the SSRL beamlines 5-2 and 5-4 with an energy resolution of 9--25\,meV depending on the beamline and photon energy used. The angular resolution is 0.1$^\circ$. The base pressure stayed below $4\cdot 10^{-11}$\,torr. We use linear horizontal (LH) and linear vertical (LV) light polarization to highlight different parts of the band structure. The samples were cleaved {\it{in-situ}} below 30\,K.


\section{Results and Discussion}

\subsection{$4f$ Electronic Structure from Photoemission Spectroscopy}

\begin{figure}
\begin{center}
\includegraphics[width=1\columnwidth]{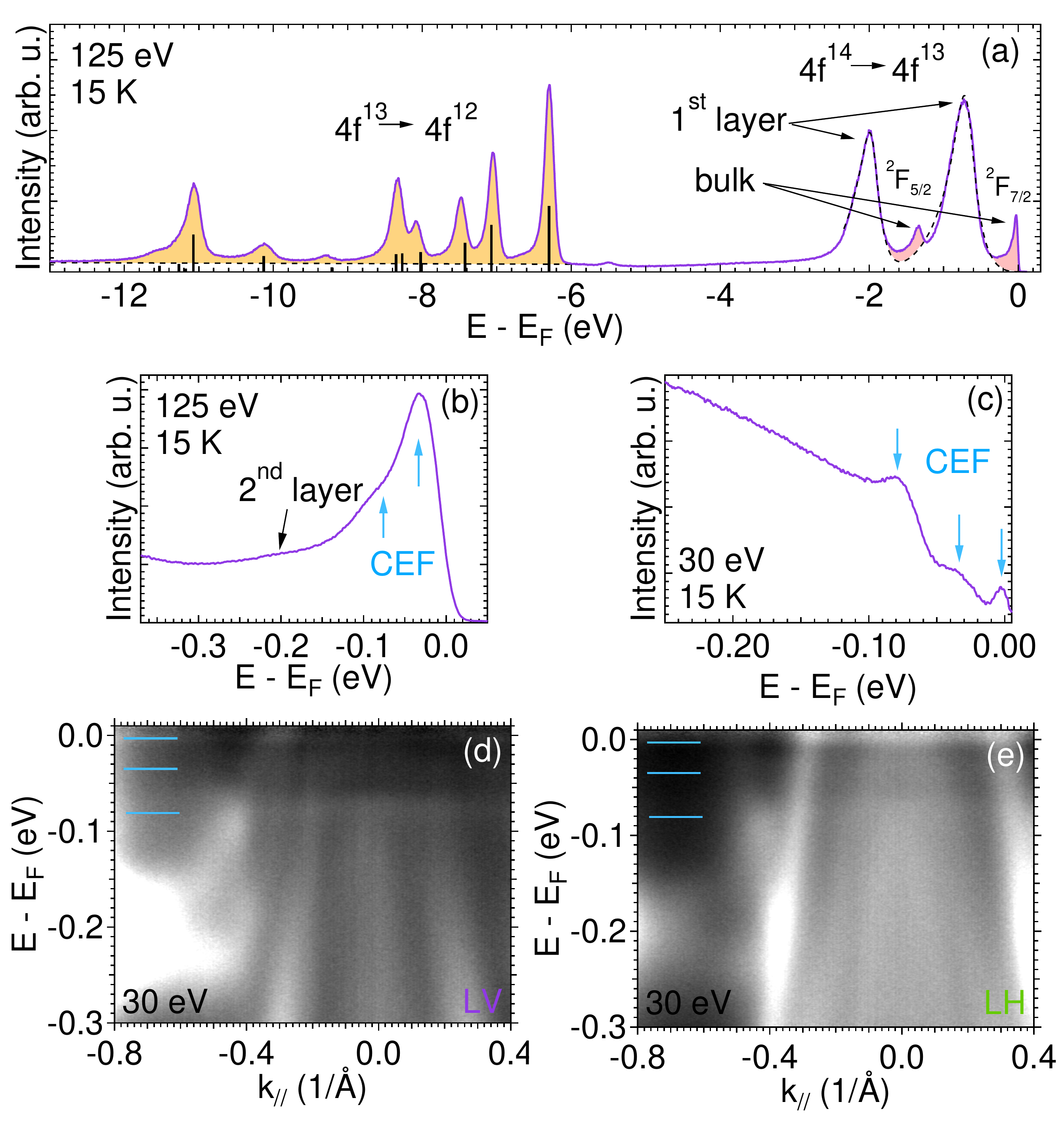}
\caption{Yb $4f$ valence band structure. (a) Integrated EDC taken at $E_{h\nu} = 125$\,eV  measured with LV polarization. The binding energies and relative intensities of the Yb$^{3+}$ multiplet ($4f^{13} \rightarrow 4f^{12}$) are well reproduced by the calculations\cite{gerken_1983} (black bars). The bulk Yb$^{2+}$ doublet ($4f^{14} \rightarrow 4f^{13}$) has two counterparts originating from the 1$^\mathrm{st}$ (surface) Yb layer and from the 2$^\mathrm{nd}$ (burried) Yb layer (b). (b),(c) Zoom into the doublet ground state for $E_{h\nu} = 125$\,eV and 30\,eV. (c) shows the EDC divided by a Fermi-Dirac distribution to highlight the CEF level close to $E_\mathrm{F}$. The arrows highlight the the three observed CEF levels and intensity from the 2$^\mathrm{nd}$ Yb layer. (d),(e) ARPES spectra devided by a Fermi-Dirac distribution for LV and LH polarization at 30\,eV. The lines mark the position of the CEF levels.}
\label{Fig:f_levels}
\end{center}
\end{figure}

First, we study the valence band structure of the Yb 4$f$ electrons. We use a photon energy of 125\,eV. It is close to the Cooper minimum of the Ru $4d$ and Yb $5d$ states \cite{yeh_1985} and therefore suppresses their contribution to the spectral weight. At the same time the photoionization cross-section for Yb $4f$ states is enhanced at this photon energy. Figure \ref{Fig:f_levels}(a) presents the photoemission spectrum at 125\,eV.

Photoemission of an electron from the 4$f^n$ configuration leads to a final state configuration 4$f^{n-1}$, which is characterized by atomic-like multiplets. Yb has a non-integer valence in heavy-fermion systems. Therefore, both Yb$^{3+}$ ($4f^{13}\rightarrow4f^{12}$) and Yb$^{2+}$  ($4f^{14}\rightarrow4f^{13}$) contribute to the photoemission spectrum. Since the valence is very close to 3+ and the ground state is almost degenerate with the $4f^{13}$ configuration, we find the multiplet corresponding to Yb$^{2+}$ at the Fermi level. The multiplet originating from Yb$^{3+}$ is separated from it by the Coulomb interaction $U$ between $f$-electrons.

We observe these two multiplets in the spectrum in Fig.~\ref{Fig:f_levels}(a). The  $4f^{13}\rightarrow4f^{12}$ multiplet between 12\,eV and 6\,eV consists of 13 levels that are split by spin-orbit coupling. The binding energies and relative intensities compare well with theoretical calculations \cite{gerken_1983}. The multiplet close to $E_\mathrm{F}$ corresponds to the transition $4f^{14}\rightarrow4f^{13}$. Spin-orbit coupling splits the final state into a doublet. The two additional, rigidly shifted peaks seen in Fig.~\ref{Fig:f_levels}(a) indicate photoemission from the surface Yb layer. It has been shown that the lower coordination of the surface atoms generally leads to a shift of the level to higher binding energies for rare earth elements \cite{schneider_1983} and similar magnitudes of shifts were observed in Yb systems before. \cite{gerken_1982,kaindl_1983,kummer_2011,vyalikh_2010_spec} \YRG~will most likely cleave between the Yb and the Ge atomic layer. This allows for two surface terminations. The large intensity of the surface doublet in Fig.~\ref{Fig:f_levels}(a) indicates, that the sample has a Yb terminated surface.\cite{danzenbaecher_2007} The absence of a similar surface Yb$^{3+}$ multiplet indicates that the surface Yb layer has a valence close to 2.\cite{danzenbaecher_2007} We find a third pair of peaks with very low intensity shifted by $\sim 0.2$\,eV (Fig.~\ref{Fig:f_levels}(b),Fig.~\ref{Fig:spectra}(a1,c1)). This shift is larger than the expected CEF splitting. Such a set of multiplets was also observed in YbRh$_2$Si$_2$ which was attributed to the next buried layer of Yb.\cite{kummer_2011} We therefore interpret the third doublet in \YRG~to photoemission from the 2$^\mathrm{nd}$ Yb layer, \ie~the first buried layer. Similar observations of multiplets from the first and the second surface layer have been found in YbAl$_3$ \cite{suga_2005}. The third doublet is not a signature of a mixed surface termination consisting of both Yb terminated and Ge terminated patches. It has been shown in \YRS, that both terminations would result in a surface doublet at the same binding energy but with very different intensities relative to the bulk doublet \cite{danzenbaecher_2007}.  In our measurements, the 4$f$ levels are generally enhanced in LV polarization for all bulk levels. The surface states in contrast are more pronounced in LH polarization. This could relate to the finite out-of-plane component of the LH polarization due to the finite incident angle of the light.

One can determine the Yb valence from the relative intensities of the Yb$^{3+}$ and the Yb$^{2+}$ multiplets.\cite{kummer_2011} From the relative peak heights we already expect the valence to be closer to 3+ in \YRG~compared to \YRS~in accordance with thermodynamic and transport measurements.\cite{jeevan_2006} A quantitative analysis requires a subtraction of surface contributions and of a background from inelastically scattered electrons as well as valence band contributions \eg~from Ru 4$d$ electrons. The contributions for the Yb$^{3+}$ multiplet are small and we estimate the background by a linear function. For the Yb$^{2+}$ doublet, the peaks from the surface layer have the largest contribution to the background. We estimate them by fits using a Doniach-Sunjic line shape convoluted by a Gaussian. The resulting intensities of the Yb$^{3+}$ and Yb$^{2+}$ multiplets are shown as shaded areas in Fig.~\ref{Fig:f_levels}(a). They include photoemission from the bulk and from the first buried Yb layer, which have almost the same valence in \YRS.\cite{kummer_2011} We find an estimate for the valence in \YRG~of 2.95(4).

Figure \ref{Fig:f_levels}(b)-(d) shows the CEF splitting at 125\,eV and 30\,eV. We highlight the ground state CEF level close to the Fermi level in Fig.~\ref{Fig:f_levels}(c-e) by dividing with the Fermi-Dirac distribution. We can resolve three CEF levels at a binding energy of $E_0=(3\pm3)$\,meV, $E_1=(35\pm3)$\,meV and $E_2=(78\pm3)$\,meV. This observation fits well to the CEF scheme described in literature. The splitting of the quasi-quartet of less than 1\,meV \cite{jeevan_2006,jeevan_diss} is below our experimental resolution and we observe a single level at $E_0$. The splittings to the next excited levels are $(32\pm6)$\,meV and $(75\pm6)$\,meV. The first agrees with the value from inelastic neutron scattering.\cite{jeevan_2011} The largest splitting was so far only estimated to 91\,meV \cite{jeevan_diss}, which is slightly larger than our experimentally determined value.

\subsection{LDA Calculations}

\begin{figure}
\begin{center}
\includegraphics[width=\columnwidth]{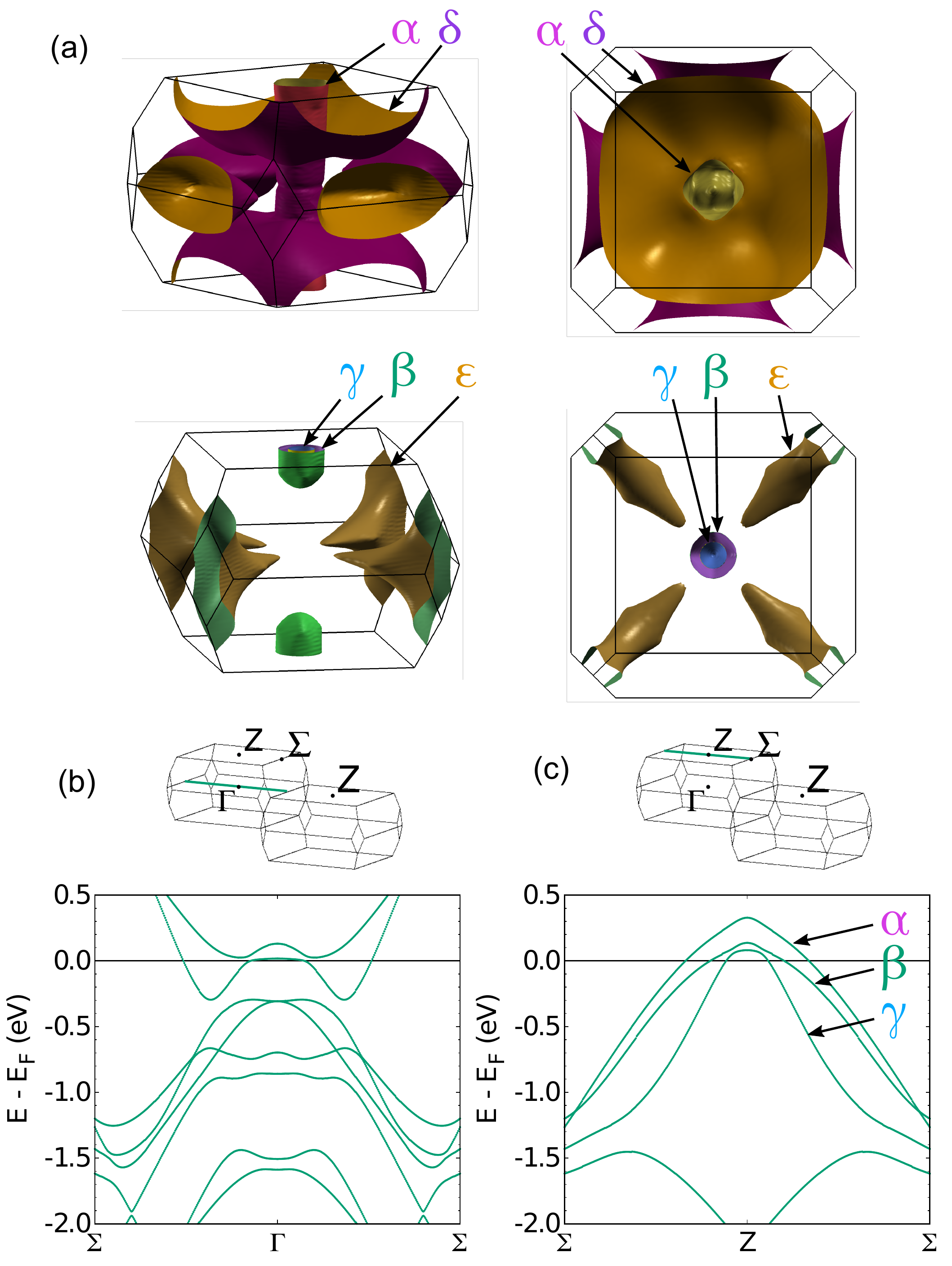}
\caption{
Calculated Fermi surfaces and band structure. The calculations are performed for LuRu$_2$Ge$_2$ using the experimentally determined lattice parameters of \YRG~ \cite{francois_1985}. They include effects of spin-orbit coupling. (a) We find five Fermi surfaces $\alpha$--$\gamma$ which are shown in two separate sets for clarity. (b) Band structure along momentum cuts as marked in the BZ. They correspond to the experimental cuts shown in Fig.~\ref{Fig:spectra}.
}
\label{Fig:FS_calc}
\end{center}
\end{figure}

We performed LDA calculations of the band structure using the wien2k package including effects of spin-orbit coupling. The calculations are performed for LuRu$_2$Ge$_2$ using the experimentally determined lattice parameters from \YRG~\cite{francois_1985}. The Lu-counterpart has a fully occupied $f$-shell removing the 4$f$ electrons from the valence band structure. The bands shown in Fig.~\ref{Fig:FS_calc} therefore originate primarily from Ru $4d$ electrons. The hybridization of these electrons with the renormalized flat 4$f$ electron bands in the Kondo lattice state of \YRG~will be clearly visible in the photoemission spectra and will only slightly alter the Fermi surfaces. 

We find five Fermi surfaces: a cylindrical almost two-dimensional sheet ($\alpha$), two small hole pockets at Z ($\beta$,$\gamma$) and two large three-dimensional sheets ($\delta$,$\epsilon$). The shape and size of $\epsilon$ strongly depends on the position of the Fermi level. Interestingly, the cylindrical $\alpha$-sheet has a square-like shape implying preferable nesting conditions.

\subsection{ARPES Results in the Tetragonal State of \YRG}


\begin{figure}
\begin{center}
\includegraphics[width=\columnwidth]{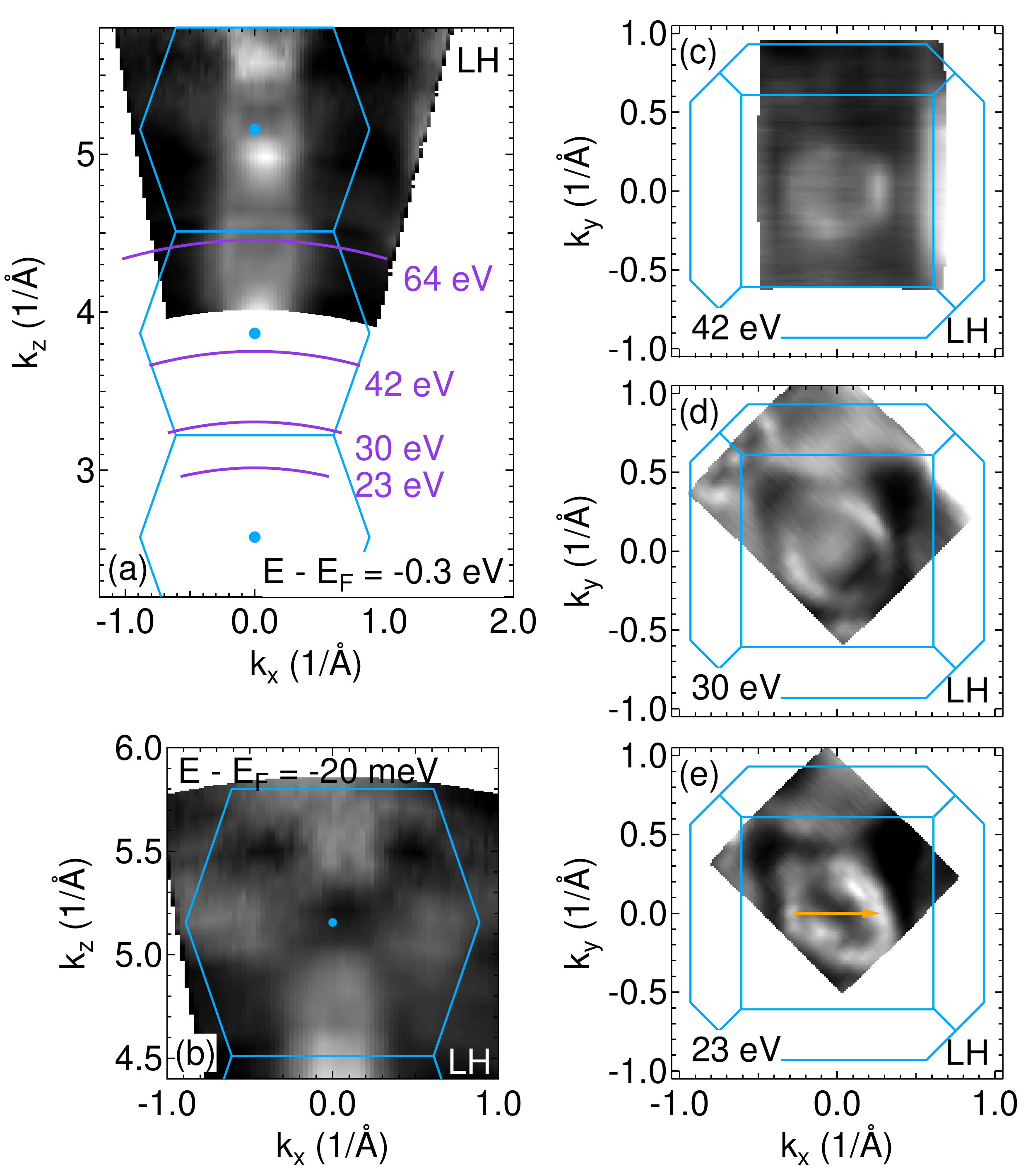} 
\caption{
Photon energy dependence and FS maps in the normal state at 15\,K. (a) and (b) show photon-energy dependencies for a cut through the zone center along $k_x$ at a binding energy of (a) 300\,meV and (b) 20\,meV. (c)-(e) FS maps for different photon energies that probe different $k_z$ as marked in (a). The yellow arrow in (e) marks the magnetic ordering wave vector \cite{jeevan_2011}. Blue lines indicate the Brillouin zone.
}
\label{Fig:maps}
\end{center}
\end{figure}

Figure \ref{Fig:maps}(a) presents a map of the photon energy $E_{h\nu}$ dependence for a cut along $k_x$ through the zone center. We find a clear periodic pattern at a binding energy of 300\,meV. As we see from Fig.~\ref{Fig:f_levels} this binding energy probes mainly conduction band electrons and has a low contribution from the f-electron states. We can determine $k_\perp$ from the periodicity and indicate the relevant photon energies used throughout this study. In Fig.~\ref{Fig:maps}(b) we show a photon energy dependence close to the Fermi level. The star-shaped pattern contains contributions from the $\alpha$ to $\delta$ Fermi surfaces. Figures \ref{Fig:maps}(c)-(e) depict Fermi surfaces measured at specific photon energies. We can clearly identify one (42\,eV) or two (30\,eV,23,\,eV) round Fermi surface pockets at the zone center which we attribute to the $\alpha$ and $\beta$ sheets. The calculations predicted that parts of the $\alpha$ Fermi surface is nested. Neutron scattering found an incommensurate magnetic ordering wave vector of $q=(0.352,0,0)$ \cite{jeevan_2011}, which can be a sign of preferable nesting conditions in \YRG. The experimentally determined diameter of the $\alpha$ Fermi surface fits well to the size of $q$ (Fig.~\ref{Fig:maps}(e)).


\begin{figure}
\begin{center}
\includegraphics[width=\columnwidth]{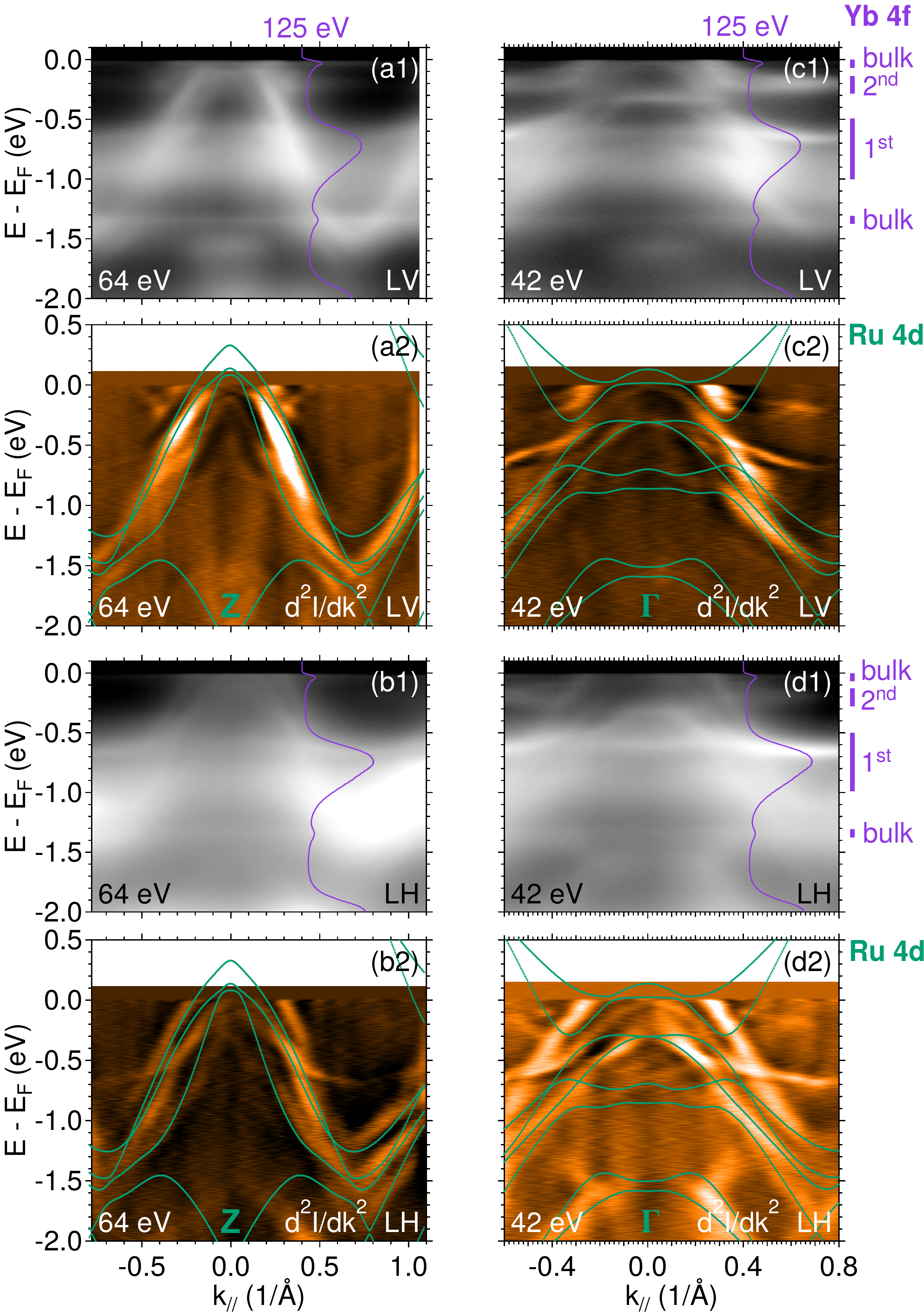}
\caption{
Spectra at $\Gamma$ and Z compared with calculations. (a1,a2) show the spectrum and its second derivative taken with 64\,eV photons in LV polarization at 15\,K. (b) Same as (a) for LH polarization. (c,d) same as (a,b) for 42\,eV. Lines in (a1)-(d1) show the integrated EDC from Fig.~\ref{Fig:f_levels}(a) taken with 125\,eV. We highlight the different bulk and surface ($1^\mathrm{st}$ and $2^\mathrm{nd}$ Yb layer) f-levels on the right side. From Fig.~\ref{Fig:maps}(a) we find that (a,b) are close to the Z-plane while (c,d) are close to the $\Gamma$-plane. We overlay the second derivative spectra with the calculated band structure from Fig.~\ref{Fig:FS_calc} that contains bands of mainly Ru $4d$ character.
}
\label{Fig:spectra}
\end{center}
\end{figure}

We now compare the calculated band structure shown in Fig.~\ref{Fig:FS_calc}(b,c) with ARPES spectra taken with corresponding photon energies of 42\,eV and 64\,eV in Fig.~\ref{Fig:spectra}. We can clearly identify the flat $4f$ bands in the ARPES spectra in Fig.~\ref{Fig:spectra}(a1)-(d1). For comparison we also added the integrated EDC from Fig.~\ref{Fig:f_levels}(a) that indicates the $f$ spectral weight. The band width of the bulk $f$-levels is much smaller compared to those from the first and second Yb layer. The $4f$ band from the second Yb layer is now clearly visible. The Ru $4d$ bands hybridize with all observed Yb $4f$ bands.

To highlight the $4d$ contribution, we take the derivative of the spectra with respect to momentum and overlay them with the calculated band structure in Fig.~\ref{Fig:spectra}(a2)-(d2). The calculated band structure fits well to our experimental results indicating that correlations among the $4d$ electrons are weak. In contrast to the calculations, the band forming the tiny $\gamma$ Fermi surface pocket is observed below the Fermi level in Fig~\ref{Fig:spectra}(a2). In compounds with a similar composition such as \YRS, clear signatures from surface related $d$ bands were observed in ARPES \cite{vyalikh_2010_spec}. Their contribution to the measured spectra depends on the surface termination. Such surface related signatures can be one reason for discrepancies between our ARPES spectra in Fig.~\ref{Fig:spectra} and the calculated bulk band structure. A detailed description of such surface related phenomena in \YRG~is necessary to perform a precise band assignment for all observed spectral signatures but goes beyond the scope of this manuscript.

\subsection{ARPES Across the Ferroquadrupole Order}


\begin{figure}
\begin{center}
\includegraphics[width=\columnwidth]{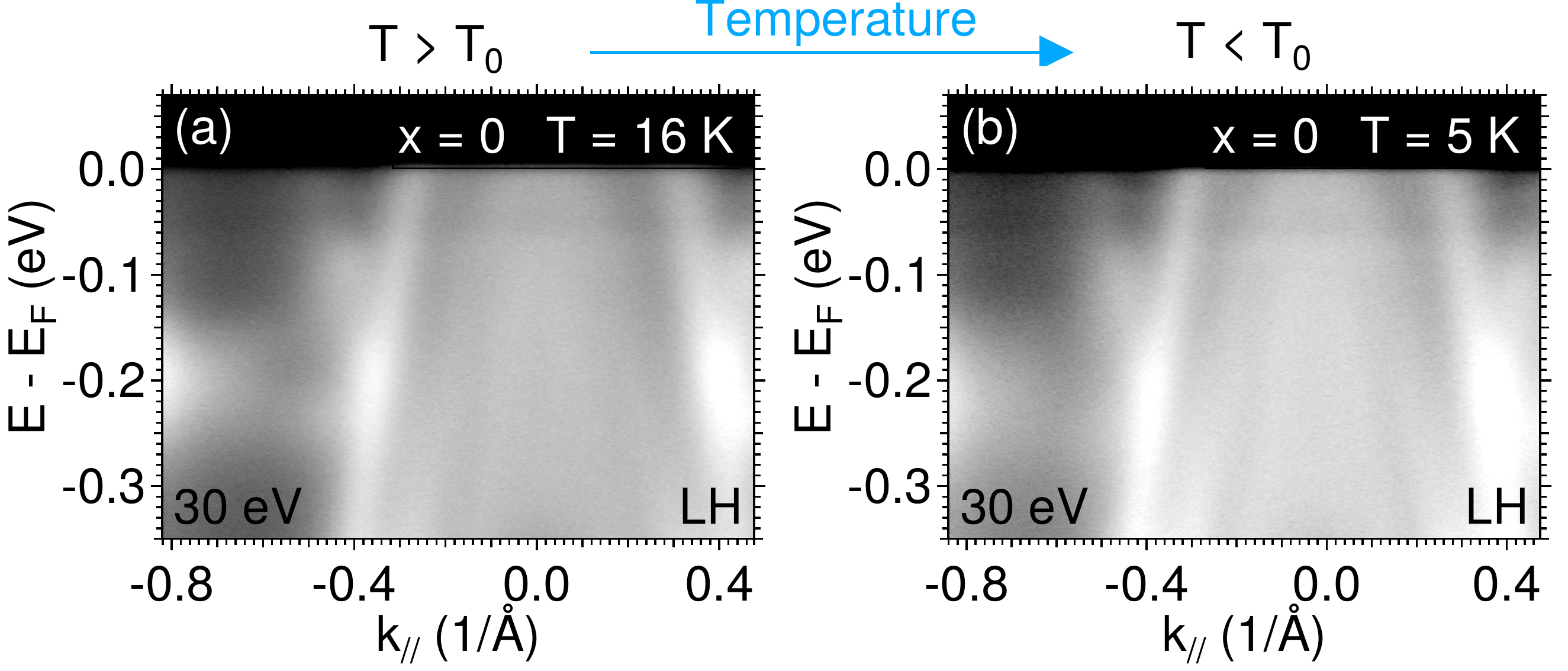}
\caption{
Temperature dependence. ARPES spectra of \YRG~along $k_x$ for a photon energy of 30\,eV which probes a $k_z$ close to Z at (a) 16\,K and (b) at 5\,K, \ie~above and below the quadrupole ordering temperature. We do not detect any changes within the resolution of our measurement.
}
\label{Fig:T}
\end{center}
\end{figure}

\YRG~undergoes a FQ order at $T_\mathrm{Q} = 10$\,K. This phase transition admixes and further splits the two closely-spaced doublets that comprise the CEF quasi-quartet ground state. Assuming a coupling between the $4f$ electrons and the conduction electrons, we also expect a two-fold distortion of the Fermi surfaces and corresponding shifts of the binding energies of the $4d$ bands. Figure \ref{Fig:T} shows a representative spectrum taken above and below the FQ order. We do not detect any change in the binding energy of the $4f$ level within our resolution. This is unsurprising as the splitting of the quasi-quartet in the normal state is already below our measurement resolution. We also do not detect any changes in the $4d$ bands. This is in contrast to the nematic phase of the FeSC, where a clear band separation between the two orthogonal directions is observed. \cite{yi_2011_pnas} For twinned crystals as studied here, this manifests in a band splitting in ARPES. Both FeSC and \YRG, however, develop an orthorhombic distortion \cite{avci_2012,rosenberg_arxiv} and a diverging nematic susceptibility \cite{chu_2012,rosenberg_arxiv} of very similar size. The transition in FeSC is driven by the $d$-electrons, whereas the ordering of the $f$-electrons are the driving force in \YRG \cite{rosenberg_arxiv}. The unchanged Ru $4d$ electronic structure in the FQ state indicates a very small coupling to the ordered $4f$ electronic system.


\begin{figure}
\begin{center}
\includegraphics[width=\columnwidth]{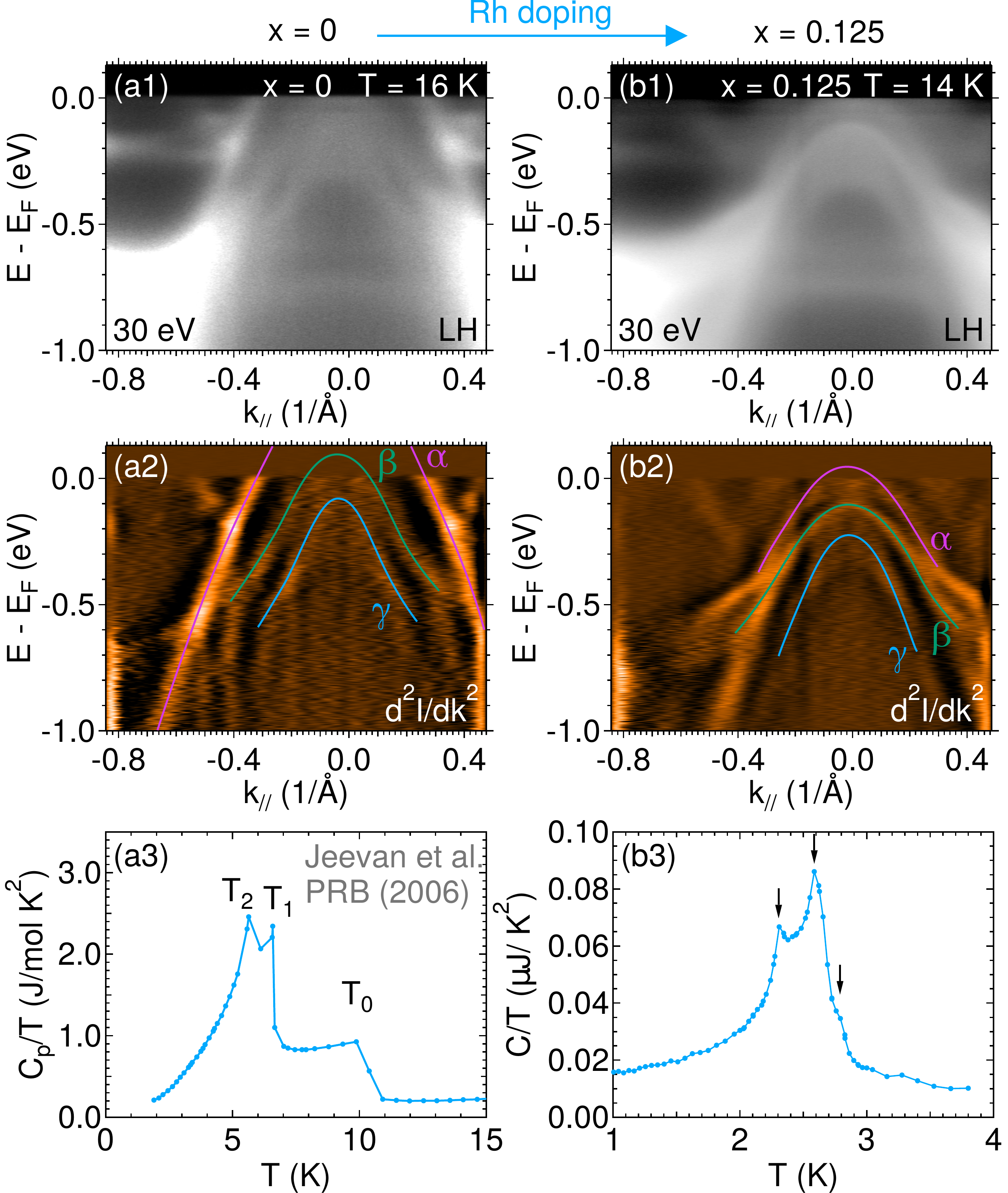}
\caption{
Doping dependence. (a1,a2) Spectrum and its second derivative for \YRG~for a photon energy of 30\,eV which probes a $k_z$ close to Z. (b) Same as (a) for Yb(Ru$_{0.875}$Rh$_{0.125}$)$_2$Ge$_2$. Lines are guide to the eye marking the position of the three hole bands $\alpha$--$\gamma$. (a3,b3) Specific heat measurements for $x=0$ reproduced from Ref.~\onlinecite{jeevan_2006} and for $x=0.125$ measured on our samples.
}
\label{Fig:doping}
\end{center}
\end{figure}

Doping \YRG~with Rh suppresses the magnetic and FQ order. We observe three phase transitions between 2\,K and 3\,K for $x=0.125$ in specific heat as shown in Fig.~\ref{Fig:doping}(a3,b3). We therefore expect Rh-doping to be a promising route to induce a magnetic and/or a FQ quantum phase transition. Figure \ref{Fig:doping}(a,b) compares two ARPES spectra for $x=0$ in the normal state and for $x=0.125$ for a photon energy that probes a $k_\bot$ close to Z. Rh-doping introduces electrons which shifts the relative Fermi level position. We indicate the positions of three hole bands in both spectra. We attribute them to the $\alpha$, $\beta$ and $\gamma$ bands that are predicted by our bandstructure calculations. The $\alpha$ and $\beta$ hole bands are shifted below the Fermi level and undergo Lifshitz transitions. The $\beta$-band vanishes at a doping level below $x=0.125$ while the $\alpha$ band likely undergoes a Lifshitz transition slightly above $x=0.125$.

The $\beta$ and $\gamma$ bands shift by approximately 150\,meV upon Rh-doping. We can compare this value to the shift predicted by the total density of states from band structure calculation. Assuming that each Rh atom contributes one additional electron and that the bands shift rigidly, we derive a shift of 90\,meV for $x=0.125$. This value is close to our measured shift.  As discussed above, we cannot exclude, that surface related signatures contribute to the spectral features in Fig.~\ref{Fig:doping}(a,b). In particular, different surface terminations can change those surface signatures. The good agreement with the calculations, however, serves as an additional support for our interpretation of a doping induced band shift.

Considering the nesting properties of the $\alpha$ band, its disappearance will likely influence the magnetic order. A change in its size can influence the ordering wave vector while its disappearance can suppress the ordering temperature.

Recent measurements of the quadrupole strain susceptibility in \YRG~indicate that the FQ order arises predominantly from magneto-elastic coupling. \cite{rosenberg_arxiv} Consequently, changes in the band filling are unlikely to have a profound effect on the FQ critical temperature $T_0$. Rather, the suppression of $T_0$ with Rh substitution is tentatively attributed to local strains induced by the chemical substitution, which act as random fields for the local quadrupoles. \cite{vojta_2013} Such an effect has previously been observed in other chemically substituted compounds which exhibit a cooperative Jahn-Teller effect similar to \YRG, for example TmVO$_4$.\cite{gehring_1976,kasten_1987}

\section{Conclusion}

In summary, we studied the heavy fermion compound \YRG~using ARPES and band structure calculations. We observe a non-integer valence of Yb and three CEF levels, which confirms the suggested quasi-quartet CEF ground state. The band structure of the conduction electrons observed in ARPES fits to our calculated band structure and we observe a hybridization with the local $4f$ levels. We do not resolve any changes in the electronic structure due to FQ order. Doping with Rh supresses the ferroquadrupolar and magnetic order. The electronic structure changes due to the electron doping and we identify two Lifshitz transitions. The Lifshitz transition close to $x=0.125$ is connected to a two-dimensional Fermi surface that is predicted to have preferable nesting properties by our calculations. Its size fits to the incommensurate wave vector of the magnetic order.

\begin{acknowledgments}

H.P. acknowledges support from the Alexander von Humboldt Foundation. E.W.R and P.M. were supported by the Gordon and Betty Moore Foundation Emergent Phenomena  in  Quantum  Systems  Initiative  through  Grant GBMF4414. This work was supported by the
Department of Energy, Office of Basic Energy Sciences, under
Contract No. DE-AC02-76SF00515. Use of the Stanford Synchrotron Radiation Lightsource, SLAC National Accelerator Laboratory, is supported by the U.S. Department of Energy, Office of Science, Office of Basic Energy Sciences under Contract No. DE-AC02-76SF00515.
\end{acknowledgments}

\bibliography{pfau_2018}

\end{document}